\documentclass[fleqn,usenatbib]{mnras}

\usepackage{newtxtext,newtxmath}

\usepackage[T1]{fontenc}

\usepackage{hyperref}

\DeclareRobustCommand{\VAN}[3]{#2}
\let\VANthebibliography\thebibliography
\def\thebibliography{\DeclareRobustCommand{\VAN}[3]{##3}\VANthebibliography}

\usepackage{graphicx}	
\usepackage{amsmath}	

\usepackage{graphics}
\usepackage{subfigure}
\usepackage{float}

\title[Galactic Emission Reflected by Meteor Trails]{The Intensity of Diffuse Galactic Emission Reflected by Meteor Trails}

\author[Zhao et al]{
Feiyu Zhao,$^{1,3}$\thanks{E-mail:zhaofy@shao.ac.cn}
Ruxi Liang,$^{2,3}$
Zepei Yang,$^{4}$
Huanyuan Shan,$^{1,3,5}$
Qian Zheng,$^{1,5}$
Qiqian Zhang,$^{2,3}$
\newauthor
and Quan Guo$^{1,5}$\thanks{E-mail:guoquan@shao.ac.cn}
\\
$^{1}$Shanghai Astronomical Observatory, Chinese Academy of Sciences, 80 Nandan Road, Shanghai 200030, China\\
$^{2}$National Astronomical Observatories, Chinese Academy of Science, 20A Datun Road, Beijing 100101, China\\
$^{3}$University of Chinese Academy of Sciences, No.1 yanqihu East Road, Beijing 101408, China\\
$^{4}$Department of Physics, Northeastern University, 360 Huntington Ave, Boston 02115, USA\\
$^{5}$Key Laboratory of Radio Astronomy and Technology, Chinese Academy of Sciences, 20A Datun Road, Beijing 100101, China
}

\date{Accepted XXX. Received YYY; in original form ZZZ}

\pubyear{2023}

\begin{document}
\label{firstpage}
\pagerange{\pageref{firstpage}--\pageref{lastpage}}
\maketitle

\begin{abstract}
 We calculate the reflection of diffuse galactic emission by meteor trails and investigate its potential relationship to Meteor Radio Afterglow (MRA). The formula to calculate the reflection of diffuse galactic emission is derived from a simplified case, assuming that the signals are mirrored by the cylindrical over-dense ionization trail of meteors. The overall observed reflection is simulated through a ray tracing algorithm together with the diffuse galactic emission modelled by the GSM sky model. We demonstrate that the spectrum of the reflected signal is broadband and follows a power law with a negative spectral index of around -1.3. The intensity of the reflected signal varies with local sidereal time and the brightness of the meteor and can reach 2000 Jy. These results agree with some previous observations of MRAs. Therefore, we think that the reflection of galactic emission by meteor trails can be a possible mechanism causing MRAs, which is worthy of further research.
\end{abstract}




\section{Introduction}

In 2014, \citet{Obenberger+2014a} found two interesting transient sources that were actually not associated with Gamma-ray bursts (GRBs) in $\sim 200$ hours of data when using LWA (Long Wavelength Array) to limit prompt radio emission from Gamma-ray bursts (GRBs). Subsequently, \citet{Obenberger+2014b} analyzed data with a total duration of 11,000 hours recorded by LWA at frequencies between 25 and 75 MHz and found 49 such transients, which are long duration (10 seconds). Ten of them were correlated both spatially and temporally with large meteors (fireballs). The ionized trails left by meteors are known to be capable of scattering radio waves such as FM broadcasting and DVB (Digital Video Broadcasting). The spectrum of reflection by meteor trails is typically narrow and strongly polarized (\citealt{Obenberger+2014b}). However, the spectrum of these sources is characterized by a broad and smooth spectrum with frequency ranging from at least 20 to 60 MHz, which is at frequencies lower than FM broadcasting and DVB and different from the narrow-band feature of reflection by the meteors. Such a phenomenon of emitting a broad and smooth spectrum by meteors is called Meteor Radio Afterglow (MRA).

\citet{Obenberger+2015} showed that the dynamic spectra of two fireballs were observed by LWA frequencies between 37 and 54 MHz, which they believed is possibly the emission of Langmuir waves. Using the Meteor Cameras Monitoring Network, they observed more radio afterglows of optical meteors and found that the amount of radio-detected meteors was strongly altitude-dependent and had a cut-off altitude of 90 km, which was consistent with the hypothesis that electromagnetic waves were emitted by electron plasma (\citealt{Obenberger+2016a}; \citealt{Obenberger+2016b}). \citet{varghese+2019} compared the spectrum of 32 MRA phenomena with 21 transmitter reflections from meteors. They concluded that the isotropic radiation pattern of MRAs is distinctly different from that of reflections. In 2020, \citet{Obenberger+2020} observed the connection between MRA and Persistent Trains (PTs) for the first time. The reaction of exothermic chemical reactions between atmospheric oxygen and ablation products drove the PTs, and the same reactions may produce the necessary suprathermal electrons to power the MRAs. \citet{Varghese+2021} fitted the spectrum of MRA and found that the distribution of the fitted slope of the spectrum peaked at - 1.73, but has no strong correlation with the physical properties of MRA.

\citet{Obenberger+2016b} also points out the possibility of observing this phenomenon with other low-frequency radio arrays. However, no events have been identified by the observation of the Murchison Widefield Array (MWA) at frequency 72-103 MHz yet, aiming to investigate meteor emissions and other transient events. In a total of 322 hours of observation, no transient candidates were identified as intrinsic emissions of meteors (\citealt{Zhang+2018}). In the observational data released by the Low-Frequency Array (LOFAR) in 2021 (\citealt{Dijkema+2021}), many MRA phenomena (hundreds of times in a meteor shower) were found in the 2020 summer and winter meteor showers (Gemini, Perseus and Quadrant). Meanwhile, LWA only found some of the brightest and far fewer, and they also found that many more MRAs were identified in summer than in winter. 

 In addition to low-frequency radio arrays, several meteor optical and radio monitoring networks have been deployed in recent years and are being expanded, such as the Belgian RAdio Meteor Stations (BRAMS) network (\citealt{2012pimo.conf...73C}), the NASA All-Sky Fireball Network (\citealt{2012pimo.conf....9C}), the Cameras for All-sky Meteor Surveillance (CAMS, \citealt{2014pim3.conf..173R}), and the Desert Fireball Network (DFN, \citealt{2012AuJES..59..177B}), which can also be used to detect the MRA. The Fireball Recovery and Interplanetary Observation Network (FRIPON), consisting of 150 cameras and 25 radio receivers deployed in Western Europe and Canada, is capable of highly automated meteorite monitoring (\citealt{2020A&A...644A..53C}). After two years of the survey, \citet{sung2020two} found no VLF (Very Low Frequency) emission signal from the fireballs. 

For now, the MRA generation mechanism remains uncertain. Such emission can be caused by a certain physical mechanism of the meteor trail itself, such as the plasma Langmuir wave (\citealt{Obenberger+2015}). It is proposed that this phenomenon may be caused by the scattering of strong signals from bright celestial radio sources by meteor trails when the Milky Way is in the sky (\citealt{Dijkema+2021}), although \citet{Obenberger+2015} argued that this possibility was low for at least three reasons. First, the reflection of emission of the Milky Way should undergo the same sporadic variations seen in reflections by meteor trails. Second, the probability that proper geometry would be present for a specular reflection of the Milky Way is very low. Third, the received power of reflections should be much dimmer than the sources they are reflecting, which is not consistent with the fact that they actually observed several cases where the received power is not much dimmer, but as high as about 10 to 20\% of the brightest possible reflected sources. The sample of MRAs used in \citet{Obenberger+2015} was limited. With more observations of MRAs (e.g. \citealt{Varghese+2021}), an estimation of the reflected signals of the diffuse emission of the Milky Way and bright celestial sources by fireballs with more details is needed to compare with the observations.   

For the reflection of meteor trails, communication engineers have carried out detailed research on this phenomenon and applied it in the field of radio communication. In the 1930s, \citet{Picard+1931} first linked the meteor shower to unexpected short-wave radio bursts. This phenomenon has been more deeply investigated after the VHF (very high frequency) band was widely used in radio and TV broadcasting. \citet{Kaiser+1952} dealt with the scattering coefficient of backward scattering of the electromagnetic waves by a column of ionization within an asymmetrical cylinder. Using a simplified approach to this problem, \citet{Eshleman+1955} determined approximate values of the reflection coefficients of electron clouds of different sizes, shapes, and densities. The total reflection of underdense clouds is calculated by summing the scattered waves from individual electrons, while overdense clouds are treated as specular reflections at the radius reaching the critical density. Hines et al. (\citealt{Hines+1956a}; \citealt{Hines+1956b}; \citealt{Hines+1957}) studied the types of distortion that may be expected for forward scattering in the path length of 1000 km in the middle latitudes, through the `cylindrical approximation of forward scattering to establish the contour charts distribution, and examined the duration of the signal, as well as the forward scattering of electromagnetic waves by overdense meteor trails, providing an effective basis for explaining and predicting the statistical effects of a large number of tracks.

In this work, we study the scattering signal from the whole sky galactic emission by overdense meteor trails to determine if the signals of MRA are caused by such phenomena. Using the existing meteor trail communication theory, the meteor trail is treated as an overdense asymmetrical cylinder trail, and we then deform the communication formula (specular reflection) to calculate the reflected signals. Furthermore, we calculate the model signals reflected by the several assumed meteor trails using the GSM (\citealt{Oliveira-Costa+etal}) sky model to estimate diffuse emission from the Milky Way and bright extended sources. In Section~\ref{sect:Method}, we describe the modelling, formula derivation process, and simulation method. The results are shown in Section~\ref{sect:result}. In Section~\ref{sect:discon}, we present the results and provide suggestions for further observational experiments. Additionally, we summarize the findings and draw conclusions in this section. The details of the formula derivation and calculation are given in the appendix.

\section{Methodology}
\label{sect:Method}

\subsection{Modelling}
After a meteoroid enters the atmosphere, the body of the meteoroid violently collides and causes friction with atoms and molecules in the air, forming a cylindrical ionization trail called the meteor trail. The degree of ionization of the meteor trail can be expressed in terms of the average number of free electrons per meter in the cylinder (electron line density). Meteoroids of different sizes produce different electron line densities when entering the atmosphere. When the electron line density is below $10^{14}$ e/m, it is called an underdense trail, while greater than $10^{14}$ e/m, it is called an overdense trail (\citealt{sugar1964radio}).
\subsubsection*{The estimation of the flux of reflected signals by meteor trails}
\label{traditional}
In previous studies of communication through meteor trails, the electromagnetic wave signal is modelled to transmit as follows: the electromagnetic wave is emitted from the transmitting station and travels in all directions; when it comes into contact with the overdense or underdense meteor trail, the signal is mirrored or scattered and travels in all directions again, and is then received by the receiving station. It should be noted that the signal is attenuated twice during this process. When the electromagnetic signal is emitted from the transmitting station, its intensity will be roughly attenuated as the inverse square of the distance from the trail because it is propagated in all directions. After the reflection of the meteor trail, it is again scattered in all directions. The pattern of attenuation depends on the shape of the meteor trail.

Based on this simplified picture, we calculate the cross-sectional flux before and after the reflection, the signal attenuation law for communication using meteor trails is derived in \citet{Hines+1957} as
\begin{subequations} \label{Hines formula}
\begin{align}
\begin{split}
P_{\rm R}\left( t \right) & =\frac{P_{\rm T}G_{\rm T}G_{\rm R}\lambda ^2\sin ^2\alpha}{32\pi ^2R_{\rm T}R_{\rm R}\left( R_{\rm T}+R_{\rm R} \right) \left( 1-\cos ^2\beta \sin ^2\phi \right)}\\ & \times \left[ \frac{4Dt+r_{0}^{2}}{\sec ^2\phi}\ln \left( \frac{r_{\rm e}q\lambda ^2\sec ^2\phi}{\pi ^2\left( 4Dt+r_{0}^{2} \right)} \right) \right] ^{1/2} ,
\end{split}
\label{Hines overdense}\\
\begin{split}
\Tilde{P}_{\rm R}\left( t \right) & =\frac{P_{\rm T}G_{\rm T}G_{\rm R}\lambda ^3q^2r_{\rm e}^{2}\sin ^2\alpha}{16\pi ^2R_{\rm T}R_{\rm R}\left( R_{\rm T}+R_{\rm R} \right) \left( 1-\cos ^2\beta \sin ^2\phi \right)} \\ & \times \exp{\left( \frac{8\pi ^2r_{0}^{2}}{\lambda ^2\sec ^2\phi} \right)}\exp{\left( \frac{-32\pi ^2Dt}{\lambda ^2\sec ^2\phi} \right)} . \label{Hines underdense}
\end{split}
\end{align}
\end{subequations}
Here (\ref{Hines overdense}) is the case of overdense trails, and (\ref{Hines underdense}) is the case of underdense trails. 
\begin{itemize}
    \item $P_{\rm T}$ is the transmitted power. 
    \item $G_{\rm T}$ and $G_{\rm R}$ are the gains from the transmitting and receiving antennas (relative to isotropic radiators) in the direction of the reflection point. 
    \item $q$ is electron line density in the trail (e/m), assumed to be uniform along the trail. 
    \item $D$ is the diffusion constant of the trails and is a function of the height of the trail ($\rm m^{2}/s$).
    \item $\lambda$ is the radio wavelength (m). 
    \item $r_{\rm e}$ is the classical radius of the electron ($\rm 2.818\times 10^{-15}\rm ~m$).
    \item $R_{\rm T}$ and $R_{\rm R}$ are the distances (m) from the reflection point (centre of the first Fresnel zone) to the transmitter and receiver, respectively.
    \item $r_0$ is the initial effective trail radius (m).
    \item $\alpha$ is the angle between the incident electric field and the wave normal from the receiver to the trail.
    \item $2\phi$ is the angle of incidence formed by the vectors $\vec{R}_{\rm T}$ and $\vec{R}_{\rm R}$. 
    \item $\beta$ is the orientation of the trail in the plane normal to the
plane of incidence containing $\vec{R}_{\rm T}$ and $\vec{R}_{\rm R}$. 
\end{itemize}
The factor outside of the fraction in (\ref{Hines overdense}) describes the variation of the trail radius $r$ with time.

There are several key features of overdense meteor trail forward scattering: the emitted signal is artificial, its frequency, intensity, and other properties are certain and come from a definite direction. 
The signal is emitted in a radial pattern by a point source, so there will be the first attenuation.
The meteor trail must be on the first Fresnel zone of the transmitting and receiving stations so that such an emission-reflection-reception process of the electromagnetic wave can occur.

\subsubsection*{The flux of reflected signals by meteor trails under local coordinate system }

\begin{figure*}
   \centering
   \includegraphics[width=\linewidth, ]{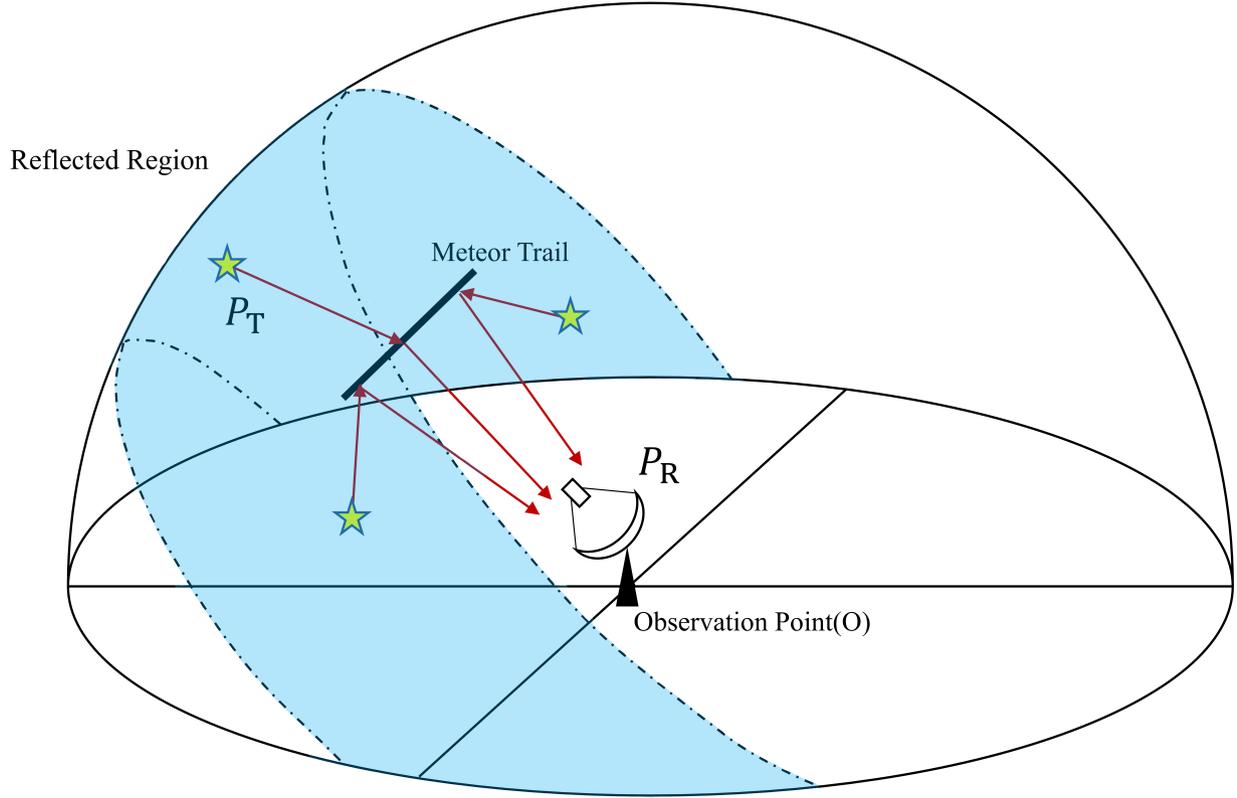}
   \caption{This figure depicts a model of radio signals from the sky that are reflected by a meteor trail. Specifically, each meteor trail has a corresponding band- or ring-shaped reflection region, as shown in the light blue region in the sky. The radio signal, represented by the yellow stars, from the blue region, can reach a receiver at an observation point after being reflected by the meteor trail. If there is a radio source or a small area with a radio emission intensity of $P_{\rm T}$ in the sky, then the signal intensity received by the antenna at the observation point after reflection from the meteor trail is denoted as $P_{\rm R}$.}
   \label{FigCon1}
\end{figure*}

The situation changes when we are dealing with the question of how meteor trails reflect signals from the sky.
The signal reflected by the meteor trails is the electromagnetic wave emitted by the sources in the sky, coming from all directions and varying in intensity at all frequencies. The electromagnetic waves emitted by astronomical sources can be considered parallel waves when they propagate to the Earth. For meteor trails at different locations, there will be different parts of the sky where the electromagnetic wave signals can be reflected to the receiving station through the trails. A conceptual sketch of the radio signal from the sky reflected by a meteor trail is given in Fig. \ref{FigCon1}.

In this study, we will mainly focus on overdense meteor trails, as the meteors (or fireballs) corresponding to the currently observed MRA phenomena are quite bright. In the overdense case, the electron concentration in the trails is beyond the threshold value for reflecting radio waves of a specific frequency and the radio waves cannot penetrate the trail. Therefore, the trail can be modelled as a metallic cylinder whose equivalent radius is the radius of the cylindrical region with a negative equivalent permittivity.

\begin{figure*}
   \centering
   \includegraphics[width=\linewidth, ]{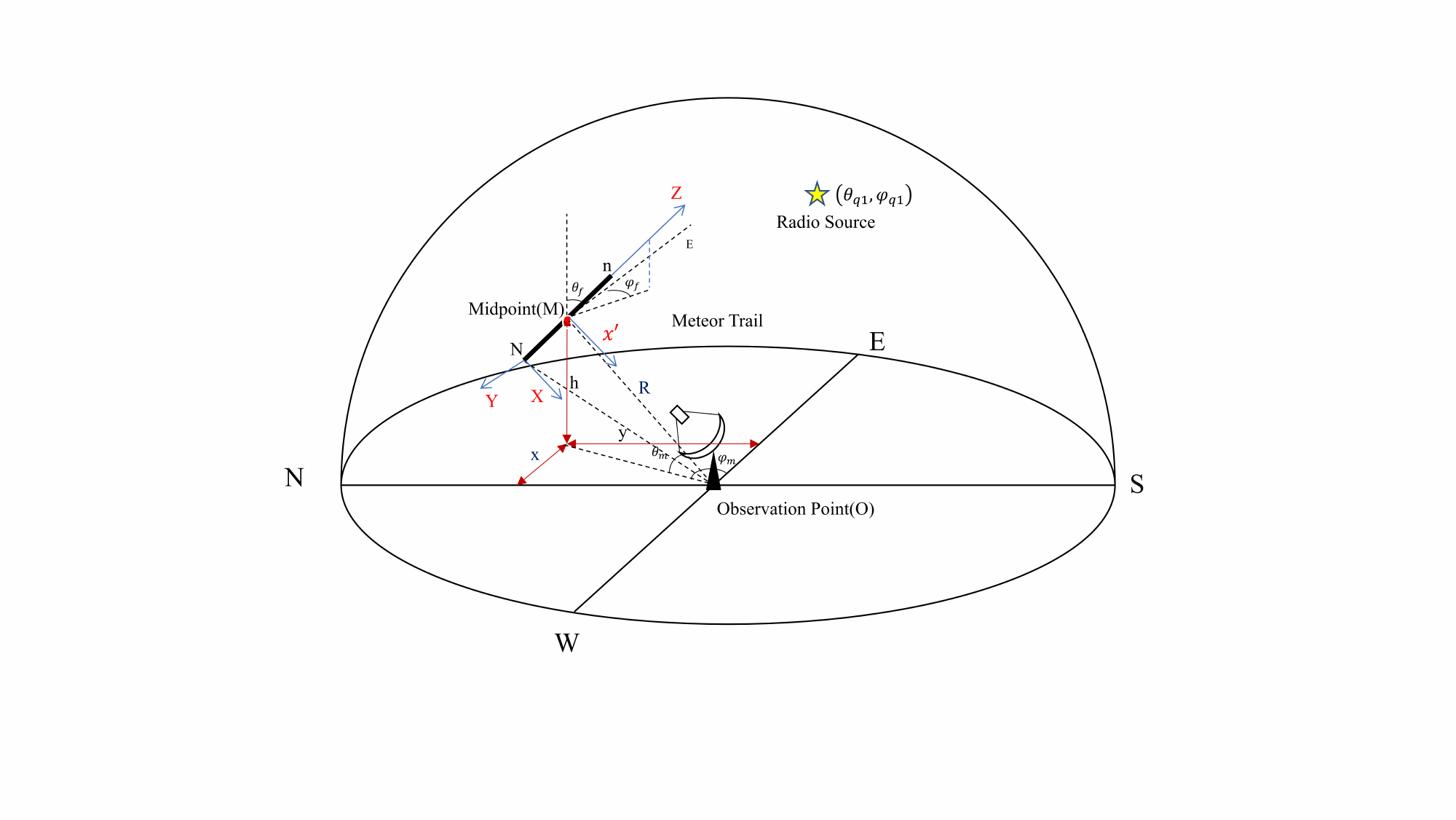}
   \caption{The figure displays the parameters utilized in the meteor-reflected sky signal model. The location of the observer or receiver is designated by the $O$ point, while the black line indicates the meteor trail. The direction vector of the meteor trail is defined by $\overrightarrow{a}$, represented by three parameters $(l, \theta_f, \varphi_f)$, and the midpoint of the meteor trail in the view of the observer is specified by its angular coordinates and distance $(R, \theta_m, \varphi_m)$. These parameters are presented in the figure. The coordinates of the radio source in the sky $(\theta_{q1}, \varphi_{q1})$ are depicted using the horizontal coordinate system, while the projection of the midpoint of the meteor trail ($M$) on the ground relative to the observation point $O$ is indicated by $x$ and $y$. Here $\left( X, Y, Z \right)$ is the coordinate system used in the ray-tracing algorithm, and $N$ and $n$ are the starting and ending points of the meteor.}
   \label{FigCon}
\end{figure*}

In the overdense case, for the incident electromagnetic wave from a certain direction, this reflection phenomenon can occur if and only if the meteor trail is tangent to an ellipsoid which has one focal on the location of the receiving station and the other focal point on the incident line. While the estimation of reflected signals equation (\ref{Hines overdense}) and (\ref{Hines underdense}) is calculated under the celestial (horizon) coordinate system, we employ another local coordinate system with the meteor trail as a reference and perform the relevant calculations mainly in this coordinate system for the convenience of calculation in the simulation later. Taking into account a meteor trail in the sky (approximated as a long cylinder) and a receiving station on the ground, the so-called local coordinate system is a right-angle or spherical coordinate system with the midpoint of the meteor central axis as the origin, the central axis of the meteor trail as the $Z$-axis and the receiving station is in the $x'OZ$ plane. Following the idea in \citet{Hines+1957}, by calculating the cross-sectional flux before and after reflection, we derived the formula for the decay of the intensity of the reflected signals by the meteor trail in a particular direction under such a local coordinate system as
\begin{subequations} \label{us formula}
\begin{align}
\begin{split}
    P_{\rm R} =  P_{\rm T}G_{\rm R}\frac{\lambda^2 }{4\pi (2 R \sin^2 \theta_q + r \sin \theta_q \cos \varphi_q)} r \cos \phi ,
\end{split}
\label{us overdense}\\
\begin{split}
    \Tilde{P}_{\rm R} =  P_{\rm T}G_{\rm R}\frac{\lambda^3 q^2 r_{\rm e}^{2}}{4\pi R  \sin^2 \theta_q } \cdot \exp\left( \frac{8\pi ^2r_{0}^{2}}{\lambda ^2\sec ^2\phi} \right)\exp\left( \frac{-32\pi ^2Dt}{\lambda ^2\sec ^2\phi} \right) .  \label{us underdense}
\end{split}
\end{align}
\end{subequations}
Here (\ref{us overdense}) is the case of overdense trails, and (\ref{us underdense}) is the case of underdense trails. 
\begin{itemize}
    \item $P_{\rm R}$ is the power received by the receiving station. 
    \item $P_{\rm T}$ is the power that can be received per unit region when the celestial signal propagates to Earth, which can be converted to flux. 
    \item $R$ is the distance from the receiving point to the centre of the meteor trail (m).
    \item $r(t)$ is the effective tail radius (m) of the meteor trail, which is a function of time, and $r_0$ is the initial effective tail radius (m).
    \item $\theta_q$ and $\varphi_q$ are the coordinates of the celestial source in the local spherical coordinate system.
\end{itemize}
The meaning of $\phi$, $q$, $D$, $\lambda$, $r_{\rm e}$ is the same as in Equation (\ref{Hines formula}).
$\theta_q$ and $\varphi_q$ are the coordinates of the celestial source in the local spherical coordinate system. It should be noted that $\theta_q$ and $\varphi_q$ are obtained by transforming based on basic parameters, and the detailed transformation formulas are given in Equation (\ref{trans}) of the appendix. The main basic parameters used in the calculation are the following:
\begin{itemize}
    \item $\overrightarrow{a}$ $\left( l,\theta _f,\varphi _f \right)$, the meteor trail direction vector, l is the half length of the meteor trail.
    \item $\left( R,\theta _m,\varphi _m \right) $, the angular coordinates and distance of the midpoint of the meteor trail in the view of the observer,
\end{itemize}
A visual illustration of these parameters is given in Fig. \ref{FigCon}.

\subsubsection*{The maximum flux of the reflected signals in the local coordinate system}
The radius of the meteor trail evolves with time. The effective tail radius (m) of the meteor trail will increase with time, while the electron line density decreases during diffusion. Therefore, the flux of the reflected signals will reach the maximum power after the diffusion starts for a period of time $t'$ (see the appendix for the details).  With the calculation of the equivalent metallic cylinder radius of the meteor trail given by \citet{14403}, we can obtain the formula of the maximum received power (see the appendix for the detailed derivation), as
\begin{equation}
\begin{split}
    P_{\rm R}  =  P_{\rm T} G_{\rm R}\frac{\lambda ^3}{4\pi^2 \left( 2 R \sin^2 \theta_q \left(\frac{e}{r_{\rm e} q}  \right)^{\frac{1}{2}}+\left( \frac{\lambda \sec \phi}{\pi} \right) \sin \theta_q \cos \varphi_q \right)} .
    \label{overdense}
\end{split}
\end{equation}
For simplicity, in this study, we assume the flux of the reflected signals is constant and equal to the maximum flux of the reflected signals described in Equation (\ref{overdense}) in this study.

We now know how to calculate the intensity of the signal from a specific position in the sky reflected towards the observation point according to Equation (\ref{overdense}), which we will use for the next step to determine which region of the sky can be reflected by the meteor trail in the next subsection. 

\subsection{Simulation}
\label{simulation}
In order to estimate the signal reflected by a meteor trail quantitatively, we run a simulation with a simplified ray tracing method to model the propagation of the reflection of the low-frequency emission of the Milky Way by meteor trails. Therefore, in the first step, we need to send a large number of rays from the observation point $O$ to the entire celestial sphere with a coarse angular spacing such as 0.1 degrees. When approaching the approximate range of azimuth and altitude of the meteor trail, a finer angular spacing, for example, 3.6', is employed.  The orientation vectors of these reflected rays after hitting the meteor trail are calculated and expressed in the horizontal coordinate system, as shown in Fig. \ref{FigCon1}.  We assume that the emission from the Milky Way propagates only along the paths of these rays.

Then, we consider the shape of meteor trails. Here, we assume that the meteor trail is a cylinder. Although the meteor trail will continue to dispute after it is generated and the trail type will change with the decrease in the electron line density during the diffusion, we assume that the flux of the reflected emission is constant, taking the maximum value of the over-dense meteor trail before it is transformed into the under-dense type, that is, $r = (\frac{r_{\rm e}q\lambda^{2}}{\pi^{2}e})^{1/2}$.

The meteor trail's effective maximal radius decreases with frequency increase. For simplicity, the effective radius used in the simulation is constant and equivalent to the effective radius at a frequency of 50 MHz.  We also assume that the centre of the bottom surface of the meteor trail cylinder is $N=[x_{0},y_{0},h_{0}]$, the angle of the cylinder's axial vector in the horizon coordinate system expressed as $\theta_{f}$ and $\phi_{f}$, and the length of the meteor trail is $L$. We can easily obtain another endpoint $n=[x_{1},y_{1},h_{1}]$ accordingly.

After we have determined the propagation of reflecting rays and the shape of the trail, we implement the method of determining whether the emitted ray intersects the meteor trail. This happens in the local coordinate system established on the meteor trail. We make the coordinate system as follows ( X, Y, Z in Fig. \ref{FigCon}):
\begin{itemize}
    \item the origin: the observation point $N$.
    \item $Z$-axis: the direction vector of the meteor trail (Z axis in Fig. \ref{FigCon}).
    \item $X$-axis: the direction of vector $NO$.
    \item $Y$-axis: $\vec{Y} = \vec{Z}\times\vec{X}$.
\end{itemize}  

 After the transformation of the coordinate system, the ray emitted from the point $O$  will be reflected by the meteor trail if:
 \begin{itemize}
     \item A projection of the incident ray on the $XOY$ plane intersects with the circle of the bottom surface of the cylinder representing the meteor trail;
     \item and then the value of $Z$-axis of the intersection point is less than the  $L$ (length of meteor trail).
 \end{itemize}
 With the intersection point,  the reflected ray will be calculated and transformed back to the horizontal coordinate system. In our ray tracing algorithm, we do not take into account the curvature of the Earth. We have verified that the impact of this approximation on the final results is in fact quite small, which is safe to be negligible in our simulations for now.

To obtain the realistic intensity of the reflected signal, we use the GSM (\citealt{Oliveira-Costa+etal}) sky model to estimate the sky temperature at a specific sky position and then convert it to the intensity. They used principal component analysis and eventually built a model with three principal components, and did not intentionally remove the point sources intentionally. It should be noted that some strong radio sources and extragalactic galaxies, such as Cas A, Cyg A, LMC, etc., are also included in the model, so these parts have been taken into account when calculating the signal intensity.  According to the received power derived in Equation (\ref{overdense}) and considering the relationship between power and flux, we can derive: 
\begin{equation}\label{2.2.2}
        I_{\rm R} =  I_{\rm T} G_{\rm R}\frac{\lambda }{2 \pi R \sin^2 \theta_q \left(\frac{e}{r_{\rm e} q}  \right)^{\frac{1}{2}}+ \lambda \sec \phi  \sin \theta_q \cos \varphi_q } .
\end{equation}

Here $I_{\rm R}$ is the flux received at the observation point by a celestial pixel that can be reflected by a meteor trail. 
The meaning of $\theta_q$, $\phi_q$, $\lambda$, $r_{\rm e}$, $e$ is the same as in Equation (\ref{overdense}).
The sum of the fluxes of all the celestial pixels that meet the conditions is the intensity of the afterglow of the meteor trail observed by the observation point radio telescope. This calculation process is represented by Equation (\ref{conv}), where $R_{\rm cof}=\dfrac{I_{\rm R}}{I_{\rm T}}$ is given by Equation (\ref{2.2.2}) and $I_{\rm T}$ is the intensity of radio emission given by the GSM model. Finally, the convolution of the reflection region determined by the ray-tracing algorithm yields the total received intensity $P_{\rm tot}$.
\begin{equation}
P_{\rm tot}=\int_{\rm reflection\ region}{R_{\rm cof}\cdot I_{\rm T}d\varOmega} .
\label{conv}
\end{equation}

\section{Result} 
\label{sect:result}

\begin{figure*}
     \centering
     \subfigure[Meteor Trail A]{
         \centering
         \includegraphics[width=0.45\linewidth]{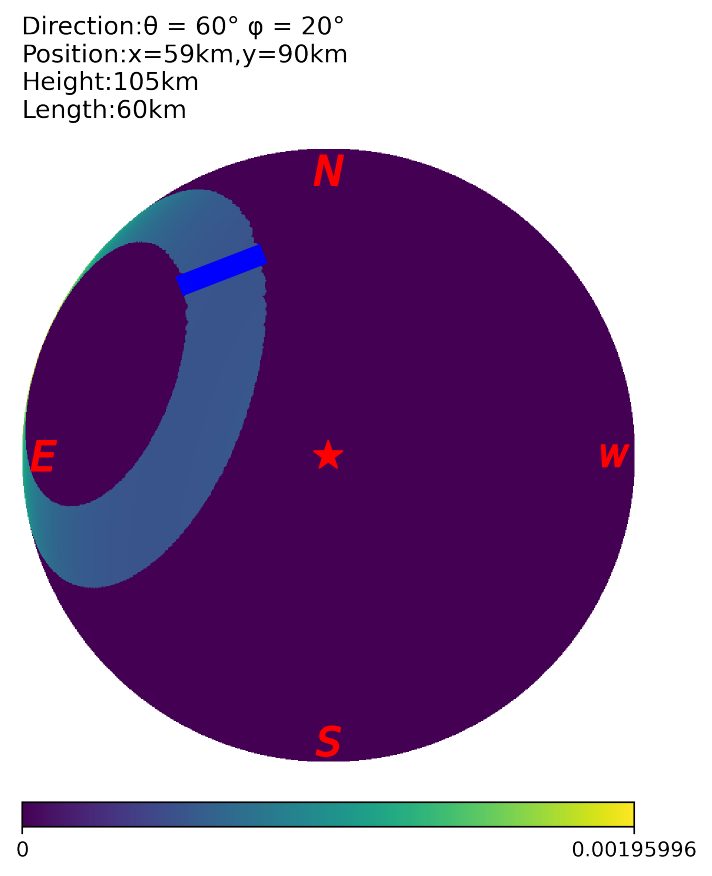}
         \label{fig:a}
         }
     \hfill
     \subfigure[Meteor Trail B]{
         \centering
         \includegraphics[width=0.45\linewidth]{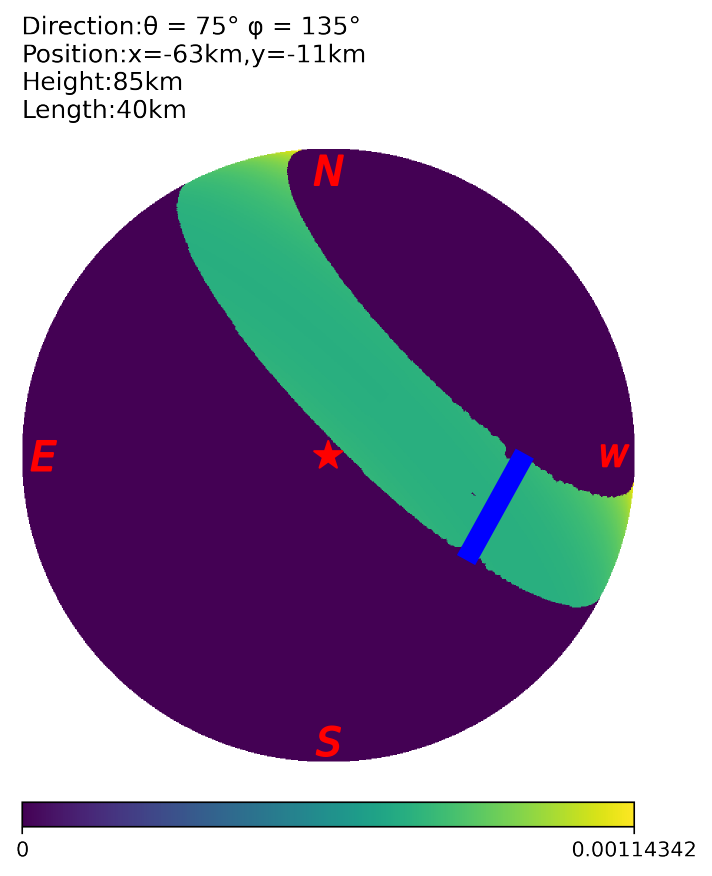}
         \label{fig:b}
         }

     \subfigure[Meteor Trail C]{
         \centering
         \includegraphics[width=0.45\linewidth]{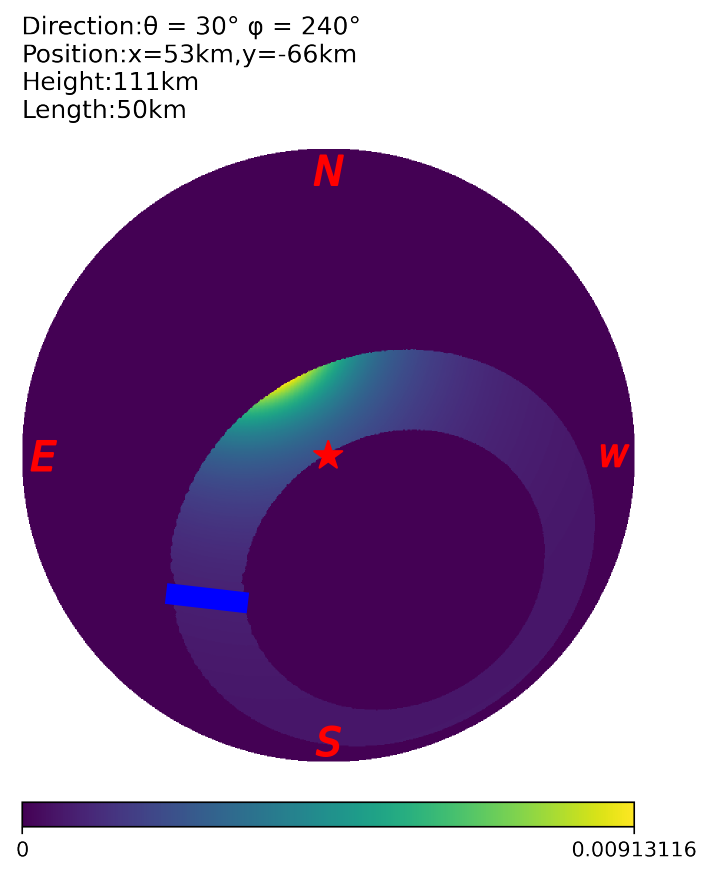}
         \label{fig:c}
         }
     \hfill
     \subfigure[Meteor Trail D]{
         \centering
         \includegraphics[width=0.45\linewidth]{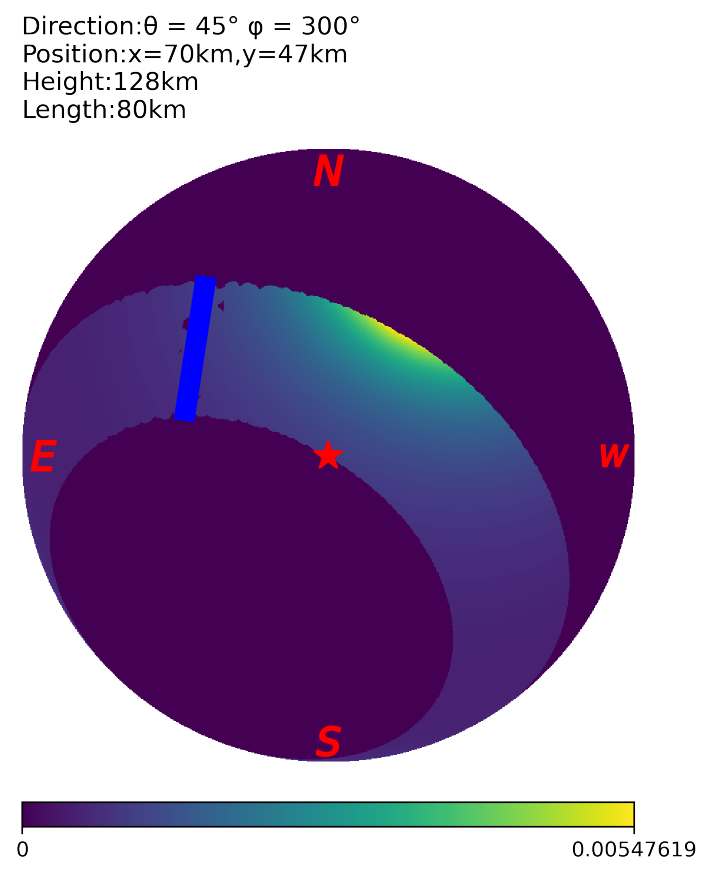}
         \label{fig:d}
         }
     
     \caption{The figure illustrates the reflection regions in the sky visible to an observer resulting from four different settings of the meteor trail parameters.  The zenith is represented by a red star, and the meteor trail is depicted as a blue line.  The bright areas in the figure indicate the reflection regions, with different colours representing various reflection coefficients at 20MHz.  The top left panel of each image presents the meteor trail parameters, where direction represents $\left( \theta _f,\varphi _f \right)$ of the meteor trail direction vector, and position and height are identical to those indicated in Fig. \ref{FigCon}. These figures use the orthographic projection.}
     \label{FigM}
\end{figure*}

\begin{figure*}
     \centering
     \subfigure[Meteor Trail E-0]{
         \centering
         \includegraphics[width=0.45\linewidth]{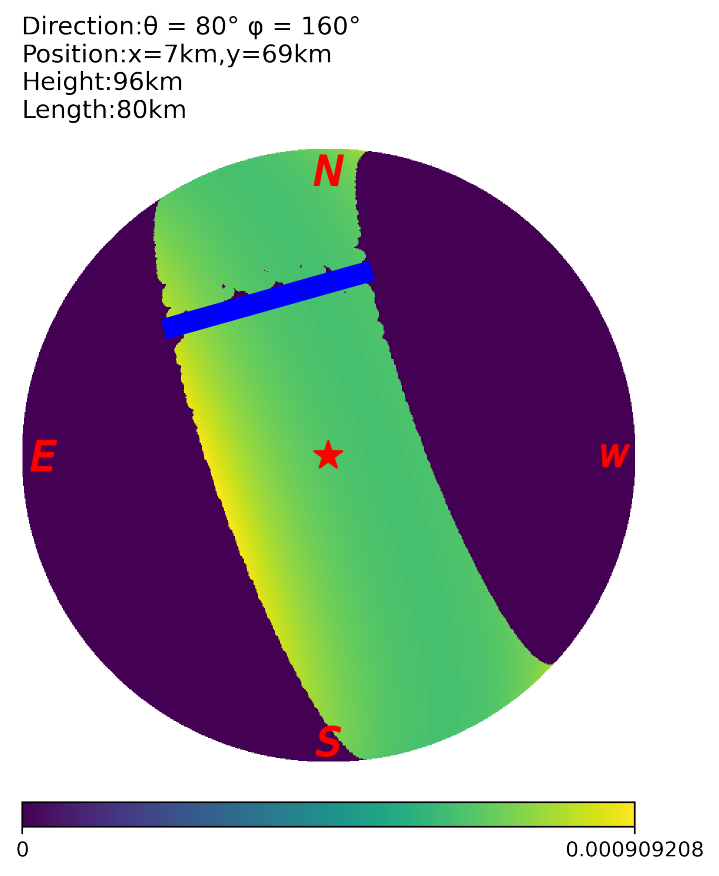}
         \label{fig:e0}
         }
     \hfill
     \subfigure[Meteor Trail E-1]{
         \centering
         \includegraphics[width=0.45\linewidth]{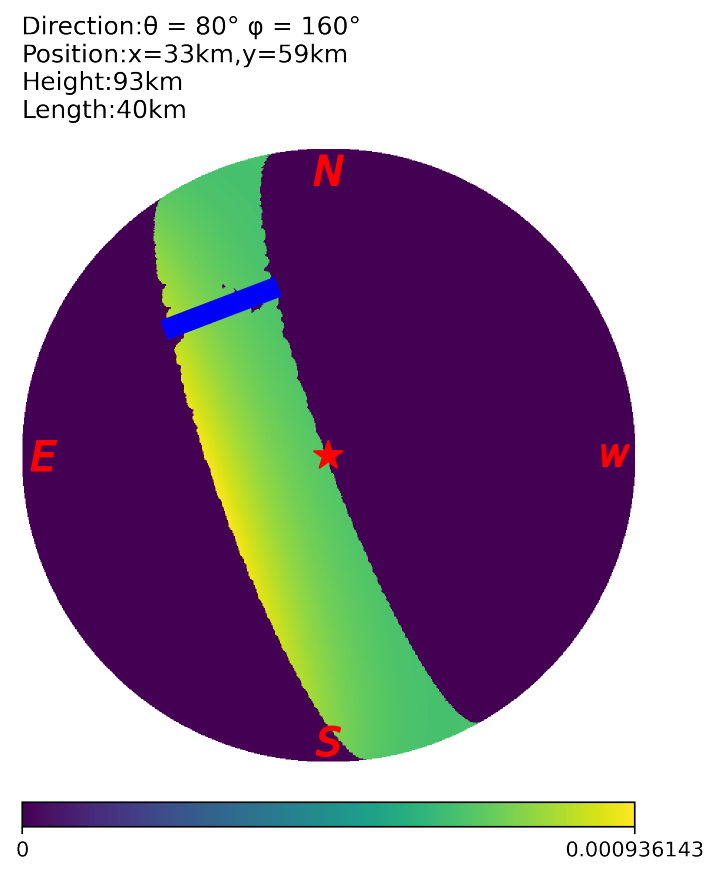}
         \label{fig:e1}
         }

     \subfigure[Meteor Trail E-2]{
         \centering
         \includegraphics[width=0.45\linewidth]{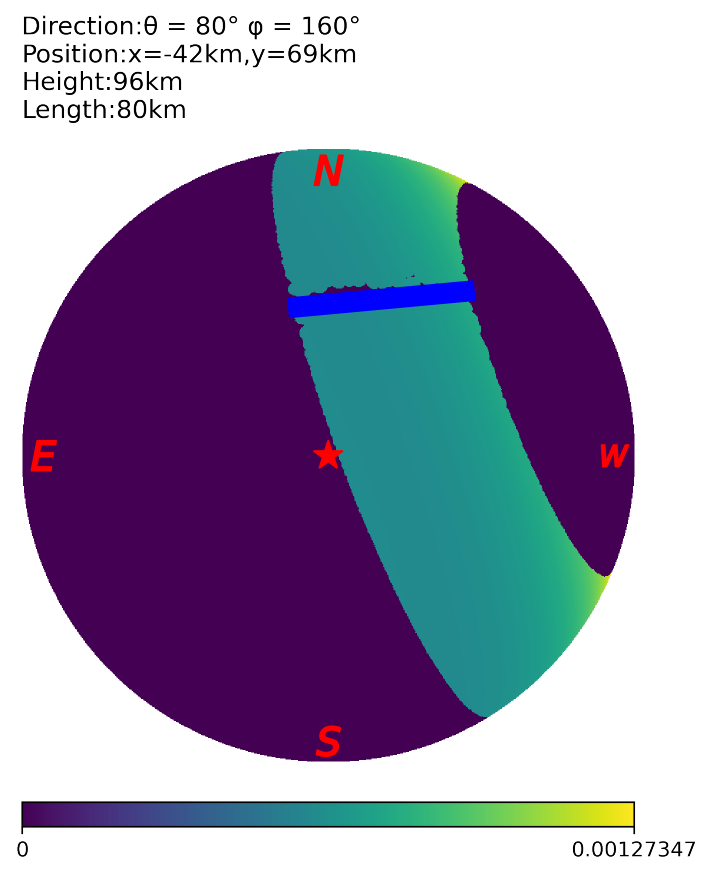}
         \label{fig:e2}
         }
     \hfill
     \subfigure[Meteor Trail E-3]{
         \centering
         \includegraphics[width=0.45\linewidth]{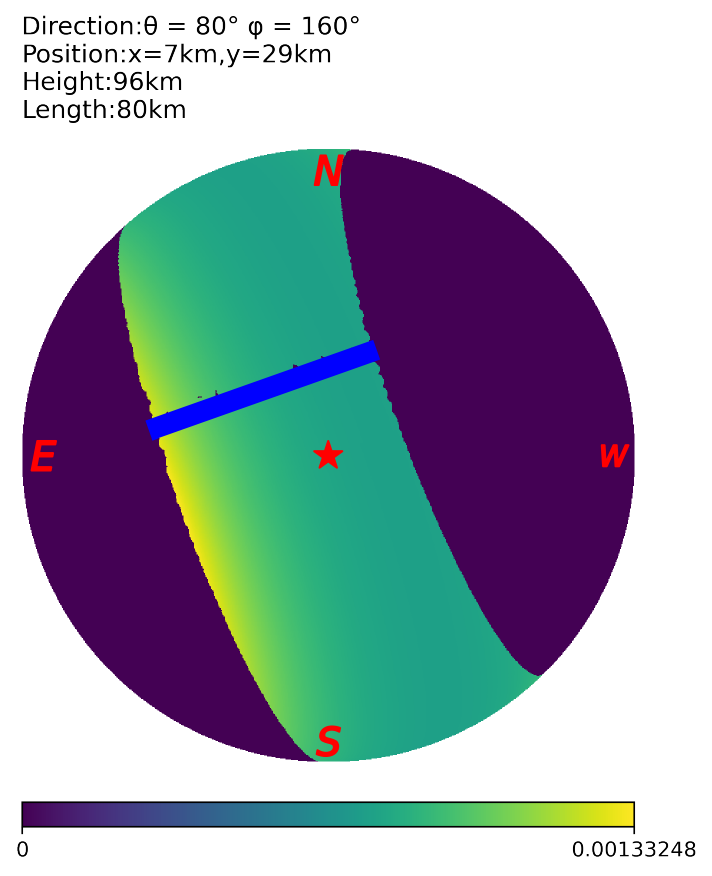}
         \label{fig:e3}
         }
     \caption{This figure shows four different meteor trails with the same direction parameters at the frequency of 20MHz. The trails are denoted as E-0, E-1, E-2, and E-3, where E-1 has a shorter length compared to the other three, while E-2 and E-3 have different position parameters. The meteor trails are represented as blue lines, and the bright regions in the sky indicate the reflection regions. The parameters for each meteor trail, such as direction parameters $\left(\theta_f, \varphi_f\right)$ and position parameters, are given in the figure. These figures use the Orthographic projection.}
     \label{FigME}
\end{figure*}

We assumed that the observer is at the site of LOFAR with latitude N52.9 ° and longitude E6.9 ° and modelled an observation of the emission of the Milky Way at a time close to the Perseids event (At 21:30 on August 15, 2022). 

In order to calculate the intensity of emission by the simulated meteor radio afterglow event, we first need to determine the region in the sky from which the signal can be received by the observer after reflection, and then calculate the reflection coefficients at the corresponding sky locations. Finally, the final intensity of the received signal is obtained by convolving the reflection coefficients at the corresponding sky locations with a frequency-specific sky emission map from the GSM model.

To determine the propagation of the reflected signals of the emission of the Milky Way by the meteor trails, we model eight cases with different positions of the meteor trails using simulation with ray tracing from Section~\ref{sect:Method}. For the parameter settings of the first four cases A B C and D, we select parameters with a large variation to represent different cases of the meteors. For the last three cases E-1, E-2, and E-3, only one of the parameters is different from case E-0 so that we can test how the results depend on the different parameters. For the first four cases out of eight, the predicted sky area from where the reflected signals can be received by the observer, together with the corresponding reflection coefficient, is shown in Fig. \ref{FigM}. The results for the last four cases are shown in Fig. \ref{FigME}. The red star in the figure represents the zenith, the blue line represents the meteor trail, and the bright regions represent the sky area that can be reflected, with colour bars indicating different reflection coefficients. The upper left corner of each panel indicates the parameters of the meteor trail used for the models. In Fig. \ref{FigME}, the direction parameters for each case are the same, except for their lengths and position parameters. The electron density of the meteor trail was set to $10^{18}e/\rm{m}$, which corresponds to a fireball of magnitude -5 with a meteoroid mass of 10 g (\citealt{Obenberger+2014a} found that the MRAs are possibly related to fireballs). The results of the models suggest that the reflection region in the sky typically appears as a band or ring determined by the projection position of the meteor trail in the observer's view. Additionally, the reflection coefficient of the reflection region varies with the length of the meteor trail. The shorter the length of the meteor trail, the narrower the reflection region. Different position parameters of the meteor trail lead to different reflection regions and reflection coefficients in the sky. E-2 and E-3 have the same meteor parameters as E-0, but are located tens of kilometres apart, while their signal intensity is about 2200 Jy, 1600 Jy, and 750 Jy at 20 MHz, respectively. From this, we can infer at least that, when observing the same event at different locations within ~100 kilometres, the intensity of the signal can vary gradually.

\begin{figure*}
     \centering
     \subfigure[Meteor Trail A, B, C, and D]{
         \centering
         \includegraphics[width=0.45\linewidth]{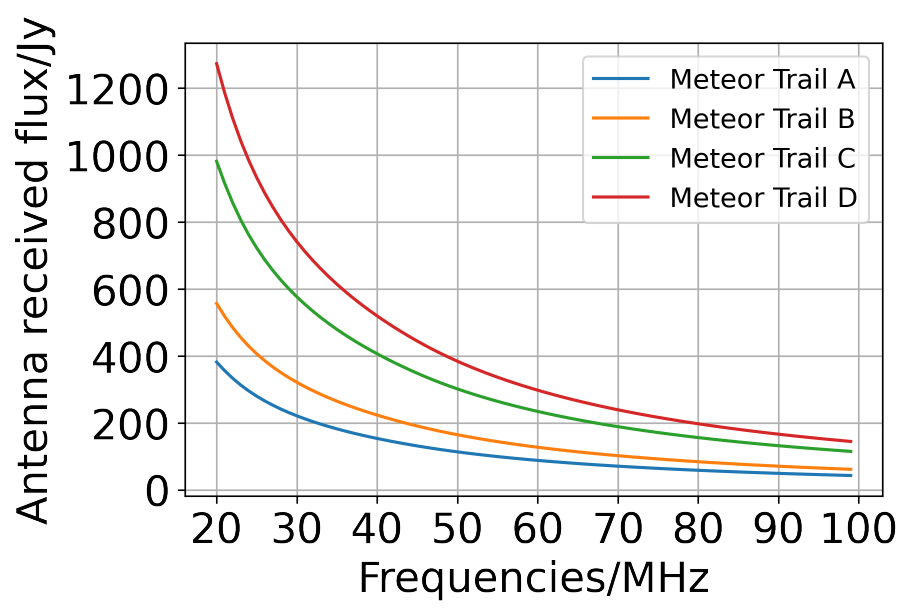}
         \label{figt:sa}
         }

     \subfigure[Meteor Trail E-0, E-1, E-2, and E-3]{
         \centering
         \includegraphics[width=0.45\linewidth]{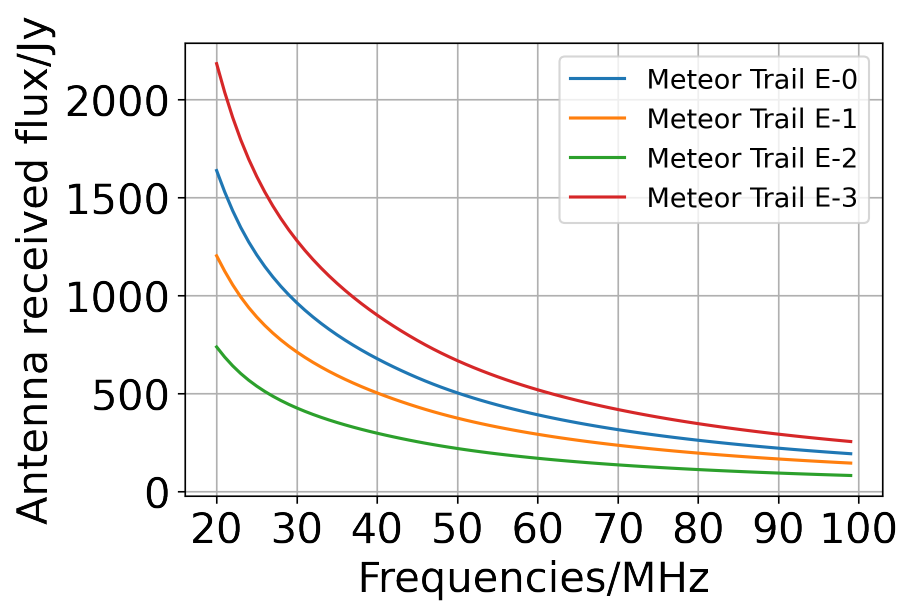}
         \label{figt:sb}
         }

     \caption{The frequency spectrum of the total received signal intensity, ranging from 20 MHz to 100 MHz at 21:30 on August 15, 2022. The horizontal axis represents the frequency in MHz, while the vertical axis indicates the total received signal intensity.}
     \label{FigS}
\end{figure*}

With the mock results of the propagation of signals, the intensity of the reflected emission from the galactic reflection by a particular meteor trail is calculated, and our next part of interest is the spectrum of meteor radio afterglow events. The observed radio sky map is generated through the Global Sky Model (GSM). We simply multiply the observed radio sky map in the frequency range of 20 - 100 MHz with the reflection coefficient of the corresponding reflection region to obtain the reflected signals by the meteor trails received by the observer at the specific location and time. The spectrum of the reflected signals of eight cases received by the observer is presented in Fig. \ref{FigS}. In these eight cases, the strongest intensity is more than 2000 Jy in 20 MHz, five times larger than the weakest one. As a reference, the signal intensity of two radio afterglow events reported by \citet{Obenberger+2016b} is about 1500 Jy and more than 3000 Jy at 22.5 MHz.

\begin{figure*}
     \centering

     \subfigure[Meteor Trail A, B, C, and D]{
         \centering
         \includegraphics[width=0.45\linewidth]{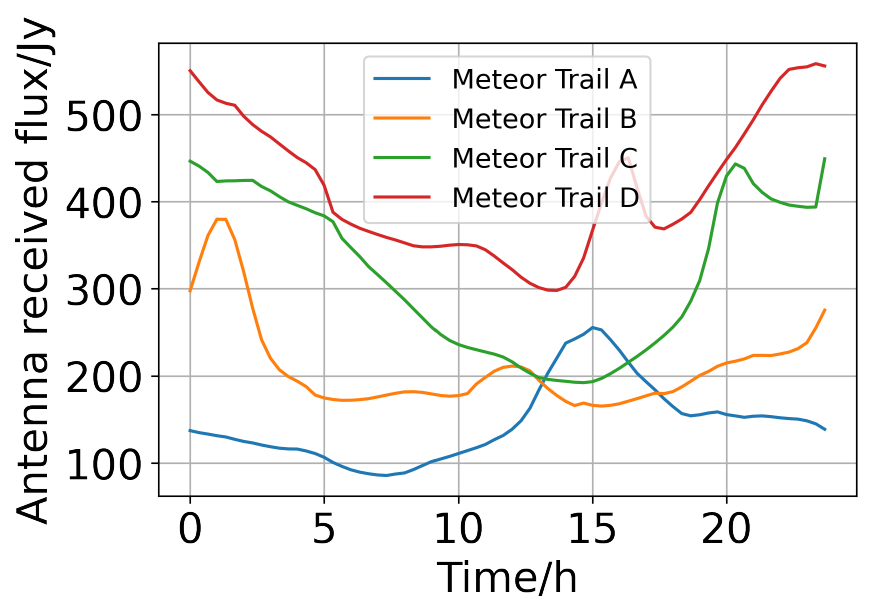}
         \label{figt:ta}
         }

     \subfigure[Meteor Trail E-0, E-1, E-2, and E-3]{
         \centering
         \includegraphics[width=0.45\linewidth]{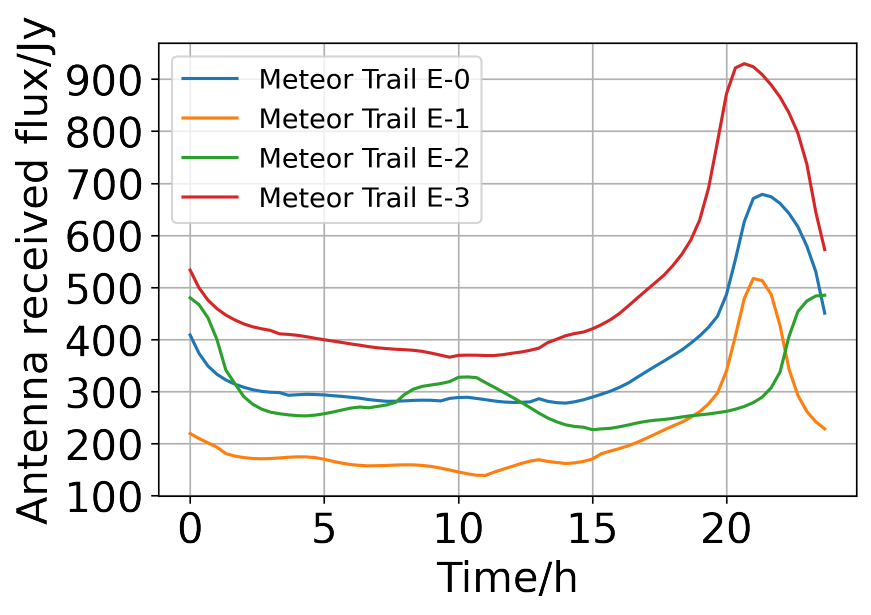}
         \label{figt:tb}
         }

     \caption{The time-dependent variation of the total received signal intensity at 40 MHz. The horizontal axis represents the time from 0:00 to 24:00 on August 15, 2022, in hours, and the vertical axis represents the received signal intensity.}
     \label{FigT}
\end{figure*}

Further, we considered how the intensity of the received reflection signal varies for the same meteor trail at different times of the day due to the rotation of the Earth, and thus the relative position of the Galactic plane. In our models, the maps of the observed radio sky are generated every 20 minutes. The total received signal as a function of the local sidereal time is calculated and shown in Fig. \ref{FigT}. Naturally, when the Milky Way is visible in the sky, the received reflection signal is stronger, and for a particular meteor trail, the intensity of the signals at different times can vary significantly. However, the reflection region of each meteor trail is distinct, and the time of the strongest signals during the day is not necessarily the same, depending on the time of strong radio sources or the Milky Way through their reflection region. The maximum and minimum values of the receiving intensity of the reflected signal for the same meteor trail can vary several times during the day.

\begin{figure*}
   \centering
   \includegraphics[width=0.45\linewidth]{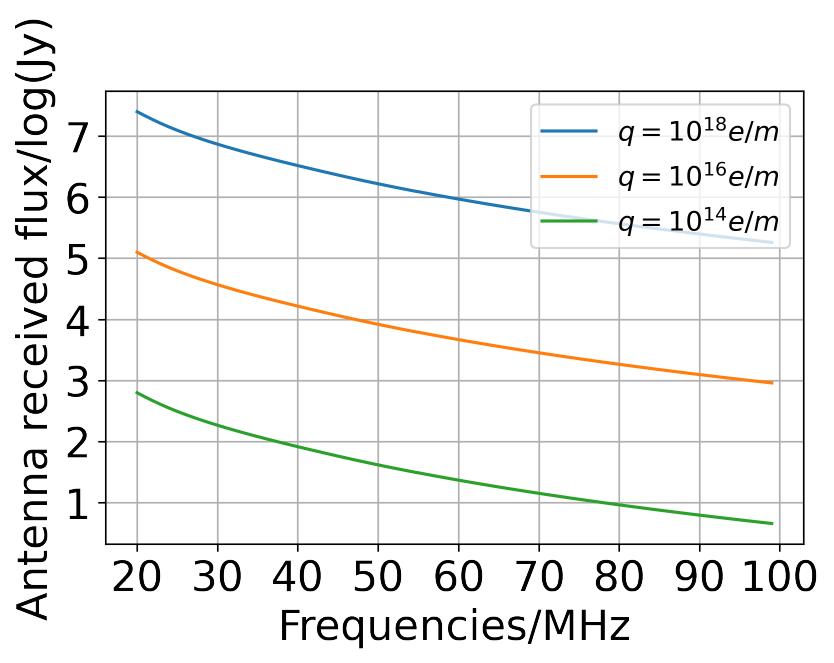}
   \caption{The frequency spectrum of Meteor Trail E-0 is presented, showing the variation in the received signal intensity as a function of frequency for different electron line densities.}
   \label{FigQ}
\end{figure*}

The brightness of meteors plays an important role in not only the intensity of the signal but also in the number of observable events.  Although bright fireball meteors can create stronger reflected signals,  the event rate of fireballs is relatively low compared to faint meteors.  We then investigate how the reflected signals depend on the meteor trails' electron line density, which is also correlated to the visual brightness of the meteors. 

\citet{jones1997theoretical} gives the analytical formula for calculating the electron line density of meteors. We select the value of the electron line density corresponding to the relation between the visual brightness and the electron line density to be calculated according to \citet{sugar1964radio}. For Meteor Trail E-0 with different electron line densities, the resulting spectrum of the reflected signals is presented in Fig. \ref{FigQ}. To further investigate the intensity of reflected signals from different meteor trails under different brightness conditions, we calculated the expected intensity of the reflection signal at 21:30 on August 15, 2022, for eight cases with different meteor brightness scenarios, and the results are in Table~\ref{tab1}. The results show that brighter meteors cause stronger reflected signals, as expected.

\begin{figure*}
   \centering
   \includegraphics[scale=0.8]{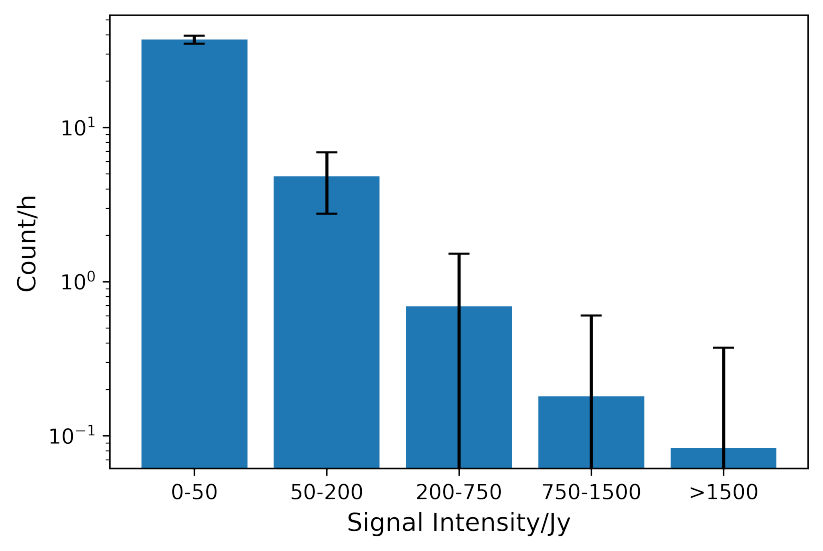}
   \caption{The intensity distribution of meteor reflection signal in an hour. We assume 43 meteors with a brightness greater than 5 magnitude per hour according to \citet{imo_vmdb} and then obtained an estimate of the distribution of the reflected signal intensity for five trails A, B, C, D, and E-0 by sampling 10000 times based on the rate of occurrence of meteors per hour of different brightness.}
   \label{Figcount}
\end{figure*}

Since \citet{Dijkema+2021} found many MRA phenomena in the meteor showers during the summer and winter of 2020 (Gemini, Perseus, Quadrant), we try to estimate the rate of occurrence of meteor radio afterglow events in a meteor shower with a toy model and to compare with the observations. We create model samples of meteors to estimate the hourly rate of occurrence of meteors of different brightnesses based on the observations of the Perseid meteor shower reported by \citet{imo_vmdb} on the night of August 12, 2022, at Ermelo,  Netherlands. Based on Miskotte's observations, an average of 43 meteors were observed per hour on that night, with meteor rates of 0.43, 0.86, 3.01, 10.11, and 28.45 for the five brightness classes given in Table~\ref{tab1}. We assume that there are 43 meteors with a brightness greater than 5 magnitudes per hour. The parameter settings except brightness are randomly drawn for cases A to E-0.  The brightness of the meteors is drawn according to the brightness distribution in Table \ref{tab1}.  We repeat the selection of model samples 10,000 times. The average and scatter of the resulting distribution of the reflected signal intensity distribution are shown in Fig. \ref{Figcount}. The average rate of the brightest reflected signals (>1500) is about 0.08 events per hour, and the rate of the faintest is about 37.2 events per hour.

 \begin{table*}
	\centering
	\caption{Expected reflected signal intensity of meteors of different brightness at 21:30 on August 15, 2022, at 20 MHz with different meteor trail settings. Different meteor brightnesses correspond to different electron line densities, and all energy intensity units are Jansky, where MB is the meteor brightness expressed in visual magnitude, and ELD is the electron line density. Assuming that the electron line density has an uncertainty of 10\%, the corresponding final received signal intensity error range is given.}
	\label{tab1}
	\begin{tabular}{lcccccccccr} 
		\hline
        MB&ELD&A&B&C&D&E-0&E-1&E-2&E-3\\
        &(e/m)&&&&&&&&\\
		\hline
        -5.0&$10^{18}$&381.59$\pm$19.10&556.78$\pm$27.87&981.05$\pm$49.11&1272.76$\pm$63.71&1637.25$\pm$81.96&1202.26$\pm$60.18&737.20$\pm$36.90&2182.53$\pm$109.23\\
        -2.5&$10^{17}$&120.66$\pm$6.04&176.09$\pm$8.81&310.30$\pm$15.53&402.57$\pm$20.14&517.79$\pm$25.91&380.24$\pm$19.03&233.12$\pm$11.67&690.24$\pm$34.55\\
        0.0&$10^{16}$&38.15$\pm$1.91&55.68$\pm$2.78&98.13$\pm$4.91&127.31$\pm$6.37&163.74$\pm$8.19&120.24$\pm$6.01&73.71$\pm$3.69&218.28$\pm$10.92\\
        2.5&$10^{15}$&12.06$\pm$0.60&17.61$\pm$0.88&31.03$\pm$1.55&40.26$\pm$2.01&51.78$\pm$2.59&38.02$\pm$1.90&23.31$\pm$1.16&69.02$\pm$3.45\\
        5.0&$10^{14}$&3.81$\pm$0.19&5.56$\pm$0.27&9.81$\pm$0.49&12.73$\pm$0.63&16.37$\pm$0.81&12.02$\pm$0.60&7.37$\pm$0.36&21.82$\pm$1.09\\
        \hline
	\end{tabular}
\end{table*}

\begin{table*}
	\centering
	\caption{The spectral index corresponds to different meteor trail settings at different meteor brightnesses at 21:30 on August 15, 2022. Where ELD refers to the electron line density, and it can be seen that the spectral index hardly changes with the change of meteor brightness for the same meteor trail setting.}
	\label{tab2}
	\begin{tabular}{lcccc} 
		\hline
        ELD(e/m)&A&B&C&D\\
		\hline
        $10^{18}$&-1.334$\pm$0.003&-1.340$\pm$0.003&-1.301$\pm$0.004&-1.319$\pm$0.004\\
        $10^{17}$&-1.334$\pm$0.003&-1.340$\pm$0.003&-1.301$\pm$0.004&-1.319$\pm$0.004\\
        $10^{16}$&-1.334$\pm$0.003&-1.340$\pm$0.003&-1.301$\pm$0.004&-1.319$\pm$0.004\\
        $10^{15}$&-1.334$\pm$0.003&-1.340$\pm$0.003&-1.301$\pm$0.004&-1.319$\pm$0.004\\
        $10^{14}$&-1.334$\pm$0.003&-1.340$\pm$0.003&-1.301$\pm$0.004&-1.319$\pm$0.004\\
        \hline
        Average&-1.334$\pm$0.003&-1.340$\pm$0.003&-1.301$\pm$0.004&-1.319$\pm$0.004\\
        \hline
        ELD(e/m)&E-0&E-1&E-2&E-3\\
        \hline
        $10^{18}$&-1.301$\pm$0.003&-1.288$\pm$0.003&-1.336$\pm$0.004&-1.306$\pm$0.003\\
        $10^{17}$&-1.301$\pm$0.003&-1.288$\pm$0.003&-1.336$\pm$0.004&-1.306$\pm$0.003\\
        $10^{16}$&-1.301$\pm$0.003&-1.288$\pm$0.003&-1.336$\pm$0.004&-1.306$\pm$0.003\\
        $10^{15}$&-1.301$\pm$0.003&-1.288$\pm$0.003&-1.336$\pm$0.004&-1.306$\pm$0.003\\
        $10^{14}$&-1.301$\pm$0.003&-1.288$\pm$0.003&-1.336$\pm$0.004&-1.306$\pm$0.003\\
        \hline
        Average&-1.301$\pm$0.003&-1.288$\pm$0.003&-1.336$\pm$0.004&-1.306$\pm$0.003\\
        \hline
        Total Average&&&&-1.316$\pm$0.003\\
        \hline
	\end{tabular}
\end{table*}

\begin{figure*}
\centering
\includegraphics[width=0.5\textwidth]{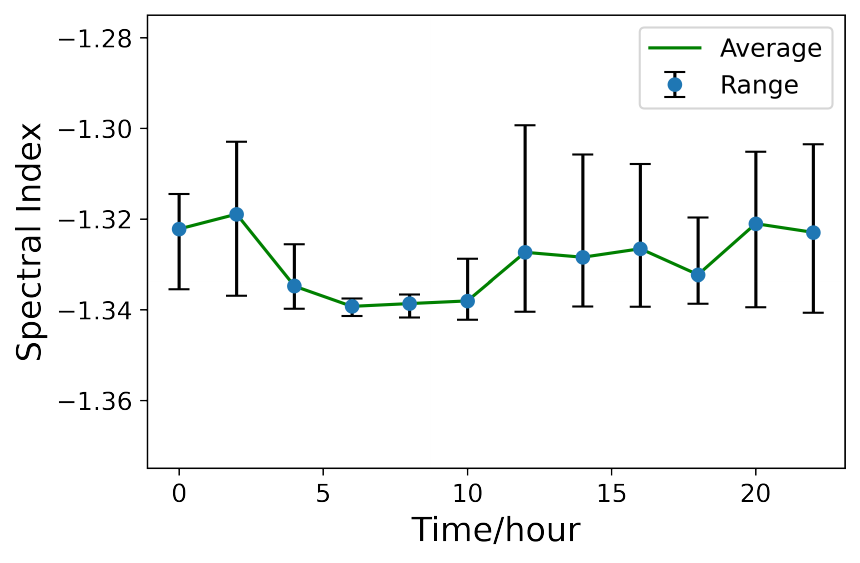}
\caption{The range of spectral index values at different brightnesses of A, B, C, D, and E-0 over time during the day. The green line indicates the average value of all spectral index, and the error bar indicates the upper and lower limits of the spectral index.}
\label{FigTS}
\end{figure*}

\begin{figure*}
\centering
\includegraphics[width=0.9\textwidth]{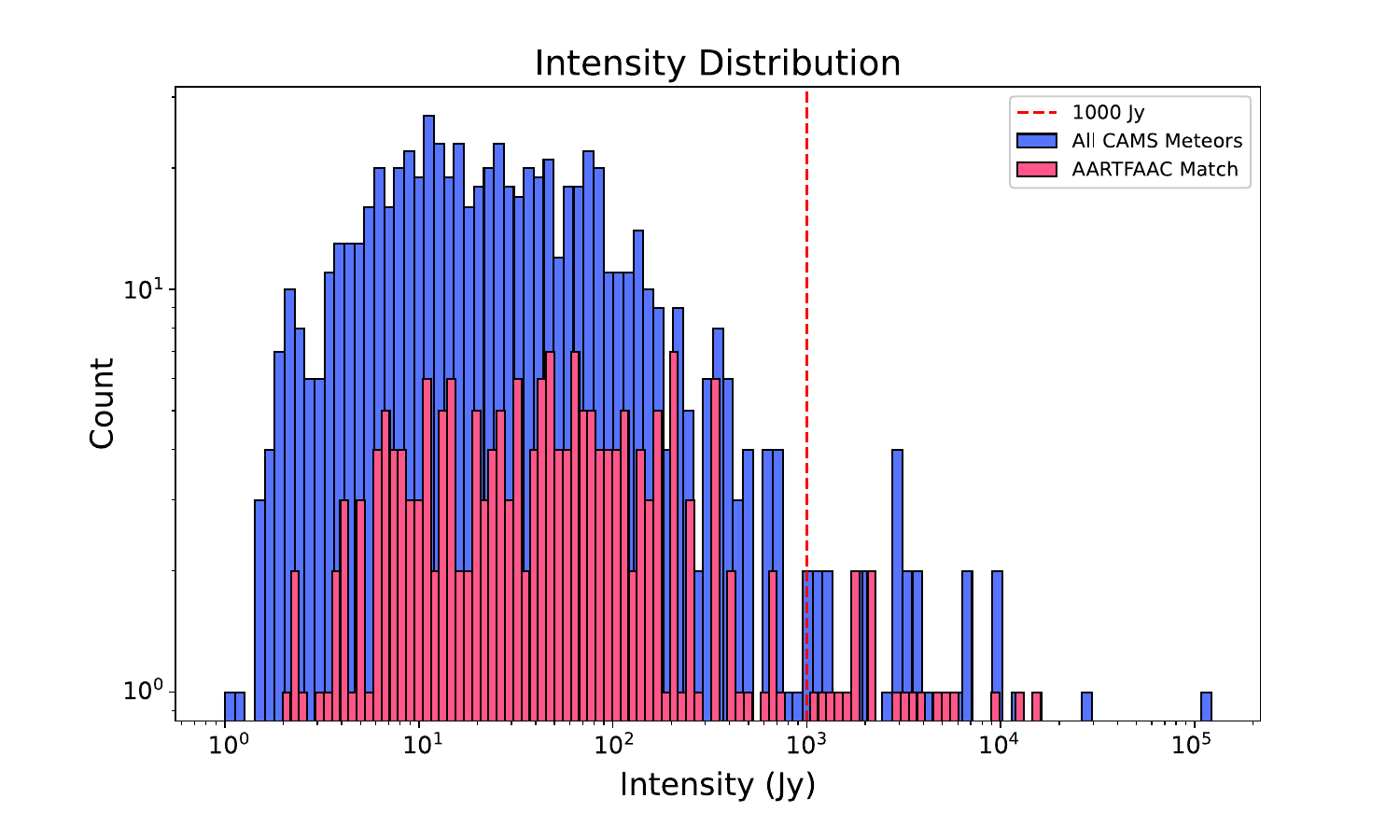}
\caption{The signal intensity distribution simulated for 721 meteor events based on the CAMS and LOFAR/AARTFAAC observations \citep{dijkema_tammo_jan_2021_5644202} at 30 MHz. The red dashed line indicates 1000 Jy, which is the typical intensity of the MRA published in \citet{Varghese+2021}. In the figure, the blue bars indicate the intensity distribution of all CAMS meteor simulations in the dataset, while the red bars indicate the simulations of meteors in the dataset with AARTFAAC radio matches.}
\label{FigID}
\end{figure*}

In previous observations, the spectral index has been recognized as a crucial parameter of Meteor Radio Afterglow (\citealt{Zhang+2018},\citealt{Obenberger+2016b},\citealt{Varghese+2021}). This study adopts a two-parameter power-law spectral equation, $I=a\cdot F^b$ as \citet{Obenberger+2016b}, to fit the spectrum from different meteor trails in our simulation. Here, $I$ and $F$ refer to the intensity of the signals and the frequency, respectively, and $a$ and $b$ denote the parameters to be determined. 

Then, we performed a power law fit to the spectra of the eight cases under different brightness conditions, and the obtained spectral index $b$ is shown in Table~\ref{tab2}. It can be seen that the spectral index of the same meteor trail hardly changes at different brightnesses, and the difference in spectral index between different meteor trails is also very small, with the largest difference only about 0.05. The average of all spectral index is -1.316201$\pm$0.003855.

To further explore the variation of the spectral index, we select A, B, C, D, and E-0 trails and fit their spectra at 2-hour intervals throughout the day using a power law, and the results obtained are shown in Fig. \ref{FigTS}. Similarly, the variation of this spectral index is small, with the maximum and minimum values differing only by less than 0.05. The spectral index varies slightly throughout the day, and we speculate that this variation may be caused by different parts of the model itself having different spectral indexes throughout the day. From the above two results, we infer that in our model the spectral index hovers around 1.316.

For comparison with actual observations, we also performed simulations based on the CAMS and LOFAR/AARTFAAC observations \citep{dijkema_tammo_jan_2021_5644202}. We calculated the intensities of the reflected signal corresponding to the 721 meteor events observed by the CAMS, and the corresponding mock intensity distributions are shown in Fig. \ref{FigID}. In this simulation, we used the meteor position, latitude, longitude, and brightness data observed by the CAMS to calculate their corresponding reflected signal intensities. Assuming the typical intensity of an MRA event 1000 Jy, our model predicts 33 MRA events brighter than this typical intensity (about 5.5 events per hour), and 19 of these events have radio matches from actual LOFAR/AARTFAAC observation. For fainter events with an intensity between 100 and 1000 Jy, our model predicts 112 events, 47 of which have radio matches from actual LOFAR/AARTFAAC observation.

\section{Discussion and Conclusion}
\label{sect:discon}
The mechanism for generating MRA is not yet clear. In this work, we calculate the reflected signals of the emission from the Milky Way by meteor trails to see if the MRA can be caused by the reflection of the strong emission of the Milky Way in low frequency.  We assume the meteor trail is an overdense asymmetrical cylinder trail and calculate the maximum flux of the reflected signals of the emission of the Milky Way. Therefore, our estimation will provide an upper limit for the reflected signals that we want to observe.  Although the ideal model utilized in this study provides valuable insight into the fundamental properties of the reflected signal, it is simplified and some physical processes are neglected. For example, the temporal structure of meteor trails that disperse over time and are subject to wind dispersion can be difficult to determine. Additionally, atmospheric conditions and ionospheric activity can also affect signal reception. Since the main purpose of the study is to determine if the reflection of the Milky Way is one of the possible reasons for MRAs, for now, we focus on the maximum flux of the reflected signals, which we compare with the observational data in other studies. Neglecting modelling the aforementioned effects will not affect the estimation of the maximum flux of the reflected signals. In our calculation, we also assume the unrealistic case that the effective radius of the meteor trails is independent of the frequency. We have verified that this simplification does not change our conclusion.

The results of our simulation indicate that the intensity of the reflected signal corresponding to a given meteor trail is subject to variation depending on the local sidereal time. The maximum signal during a day can be five times larger than the minimum signals, resulting in a higher observation rate of signals when the Milky Way is visible in the sky. This may explain the excessive number of signals detected by LOFAR during the Perseids compared to the Geminids. Furthermore, the intensity of the reflected signal from the same meteor trail sometimes gradually varies when observers are at different locations, which means that for the same event, observers at different locations not far from each other (less than 100 kilometres) may receive similar intensity of the reflected signals. 

Over several years, Obenberger et al. (\citealt{Obenberger+2014a}; \citealt{Obenberger+2014b}; \citealt{Obenberger+2015}; \citealt{Obenberger+2016a}; \citealt{Obenberger+2016b}; \citealt{Obenberger+2020}) have conducted a remarkable series of observations and studies, demonstrating that the typical maximum intensity of meteor radio afterglow lies within the range of a few hundred to a few thousand Janskys. This observation is consistent with our simulation results. Furthermore, the statistical study conducted by \citet{Varghese+2021} on the spectral index of meteor radio afterglow reveals that the index is predominantly distributed between -1 and -5, with an average value of -1.73. In our model, the spectral index approximates -1.316201$\pm$0.003855, which is roughly consistent with the observed spectral index. The spectral index in our simulation varies little with changes in the position, length, or orientation of the meteor trails, brightness, and local time of the meteors.  However, for observed cases where the spectral index is considerably larger, such as over -2 or even over -6, our model is unable to explain those cases.

\citet{Dijkema+2021} postulate that local sidereal time may be a critical factor influencing the number of meteor radio afterglow events recorded during different meteor showers. Our simulation results demonstrate that the brightness of the
reflected signals for the identical meteor trails varies at different local sidereal times, which supports this hypothesis. \citet{Dijkema+2021} also observed more faint events compared to previous studies, which is consistent with our model's prediction that fainter meteors can also generate reflected signals, albeit the flux of the signals is weaker than that produced by brighter meteors. Given that the number of faint meteors is much larger than that of bright meteors, we can anticipate a significantly larger number of faint meteor radio afterglow events, which is in agreement with the findings of \citet{Dijkema+2021}. In Fig. \ref{Figcount}, based on a toy model, we estimate that in a meteor shower, there will be 0.08 events of bright MRA (> 1500 Jy) per hour and 37 faint MRAs (0 - 50 Jy) per hour. We can compare this prediction with future observations of MRAs during the meteor shower, which will be another way to test the relation between MRAs and the reflection of the Milky Way emission by the meteor trails. 

With existing optical meteor monitoring networks, such as the CAMS and the DFN, one can determine the position, altitude, direction, and other parameters of a meteor trail. With this information, we can predict the reflected signal, which can be used to compare directly with the observation. In Fig. \ref{FigID}, we show the simulation results of the signal intensity distribution based on the CAMS and LOFAR/AARTFAAC observations \citep{dijkema_tammo_jan_2021_5644202}. 721 meteors were observed in the optical band in total, and AARTFAAC found 204 radio matches of the meteors. Taking the same parameters as the CAMS optical observation, our simulations predict 33 bright MRA (brighter than 1000 Jy) events in six hours. 19 of them actually have radio matches. It should be noted that how many of the 204 matches in \citet{dijkema_tammo_jan_2021_5644202} have board radio emission like MRA is not explicitly specified in the paper. Some events can be forward-scattering events, so a more careful comparison is needed before a conclusion can be drawn. \citet{dijkema_tammo_jan_2021_5644202} also pointed out that when the observation of the Perseid meteor was made, the Galactic centre and the Galactic plane were above the LOFAR horizon. Comparisons based on signal intensity from radio observations and meteor shower data from different seasons will help further test the model.

In summary, to explore whether the MRA can be caused by the reflection of the low-frequency emission from the Milky Way by the meteor trails,  we have derived the calculation formula for the diffuse galactic emission reflected by the meteor trail and simulated the reflection phenomenon with emission of the Milky Way modelled by the GSM. Based on the results of our simulation, we have drawn several conclusions as follows:

\begin{itemize}
    \item Our simulated signal intensity at 20 MHz can exceed a maximum of 2000 Jy, which is comparable to previous observations. In the eight cases simulated, the intensity of the maximum reflected signals can be 5 times higher than that of the minimum reflected signals. Additionally, we have found that the reflected signal is generally stronger when the Milky Way and strong radio sources are in the sky. That implies that more meteor radio afterglow events may be observed when the Milky Way is high overhead at the same observation sensitivity.

    \item We find that for the same parameter setting of the meteor trails, the different brightness of the meteors has almost no effect on the simulated spectral index, and the maximum difference in the spectral index between different meteor trails is only about 0.05, while the average spectral index for all simulated spectra is -1.316201$\pm$0.003855, which can explain part of the observed spectrum of the MRAs, but not for the cases with spectrum index steeper than -2. Perhaps there are other reasons that can also cause the MRAs. More observations and studies are still required.

    \item Since there are more faint meteors than the brighter ones,  the number of faint MRAs caused by the reflection of the emission of the Milky Way can be greater than the bright ones, for example, in a meteor shower. If the distributions of the MRA radio and the optical luminosity function are both well measured, the comparison of the prediction by our simulation with the observations may be used to study the idea that the MRAs can be caused by the reflection of emission of the Milky Way further. 

\end{itemize}

\section*{Acknowledgements}
We acknowledge the support of the Ministry of Science and Technology of China (grant No. 2020SKA0110200). HYS and QZ acknowledge the support from the Ministry of Science and Technology of China (grant No. 2020SKA0110100) and the National Science Foundation of China (11973069, 11973070). HYS acknowledges the support from the Key Research Program of Frontier Sciences, CAS, Grant No. ZDBS-LY-7013.

\section*{DATA AVAILABILITY}
The data underlying this article are available in \citet{meteorcal}, at \href{https://doi.org/10.5281/zenodo.8358749}{https://doi.org/10.5281/zenodo.8358749}.



\bibliographystyle{mnras}
\bibliography{MRA} 





\appendix
\section*{Appendix: Details of the calculation of the flux of reflected signals under the local coordinate system}

As we mentioned in the main text, in addition to the traditional celestial (horizon) coordinate system, we introduce another local coordinate system with the meteor trail as a reference and perform the relevant calculations (e.g. determining the intersection) mainly in this coordinate system. In this context, the local coordinate system is slightly different from the coordinate system discussed in Section \ref{simulation}, primarily in terms of the origin of the coordinate system. Considering a meteor trail in the sky (approximated as a long cylinder) and a receiving station on the ground, the so-called local coordinate system is a right-angle or spherical coordinate system with the midpoint of the meteor's central axis as the origin, the central axis of the meteor trail as the $Z$-axis and the receiving station lies in the $x'OZ$ plane.

Let the coordinates of a point $Q$ on the celestial sphere in the celestial coordinate system be $(\theta_{q1}, \varphi_{q1})$, where $\theta_{q1}$ is the zenith distance of the object (i.e., $\pi/2$ minus the altitude angle) and $\varphi_{q1}$ is the geodetic longitude of the object. For a meteor trail $\vec{a}$, let the midpoint of its lower base be $M$ and the focus of its upper base be $M_0$, $M$ that represents the position of the meteor trail, $\vec{a}=\vec{MM_0}$. Let the orientation of $\vec{a}$ be $(\theta_f, \varphi_f)$ in
this celestial coordinate system, the pointing of the line $\vec{b}$ from the origin to the midpoint of the meteor's central axis $M$ be $(\theta_m, \varphi_m)$, and the distance from the observatory to the meteor $|\vec{b}|=|OM|=R$, $\theta_{q1},\ \theta_m,\theta_f$ all belong to $[0,\pi/2]$ and $\varphi_{q1},\varphi_m,\varphi_f$ all belong to $[0,2\pi]$.

Then, we can transform the coordinates into the coordinates of the `local coordinate system' by three rotations of the celestial (geodetic) coordinate system. Each rotation is equivalent to multiplying each coordinate in the system by an orthogonal matrix, which preserves the distance and angle and only changes the direction of the vector, which is the same for the local coordinate system used in Section~\ref{simulation}.

It is easy to see that the zenith distance of $\vec{a}$ is $0$ after rotation.
The derivation of the coordinate transformation shows that, computing
\begin{equation}
    \label{trans}
    \begin{split}
    \theta_q & = \arccos (\sin \theta_{q1} \cos (\varphi_f - \varphi_{q1}) \sin \theta_f +  \cos \theta_{q1} \cos \theta _f) ,\\
    \varphi_q & =  \arctan \frac{\sin \theta_{q1} \cos (\varphi_f - \varphi_{q1}) \cos \theta_f -  \cos \theta_{q1} \sin \theta _f}{\sin \theta_{q1} \sin (\varphi_f - \varphi_{q1})}\\ &- \arctan \frac{	- \sin \theta_m \sin (\varphi_m -\varphi_f) \cos \theta_f + \cos \theta_m \sin \theta _f}{\sin \theta_m \sin (\varphi_m -\varphi_f)} .
    \end{split}
\end{equation}
gives us coordinate $(\theta_q,\varphi_q)$ of $Q$ in local coordinate system.

In deriving the attenuation coefficient, we followed the method of \citet{Hines+1957}, assuming a parallel light beam incident on a cylindrical trail, considering how much of the incident electromagnetic wave will be reflected. By taking a very small triangular surface element on the cylindrical surface and considering its three vertices, we can calculate the incident and reflected vectors with each point as the reflection point. Let the distance parameter of the three reflection vectors be $R$, we can obtain the region covered by the electromagnetic wave after it reflects by the trail at the distance $R$. The ratio between the region of the triangular surface element and the region after reflection is the ratio of the flux, so we can calculate the signal attenuation coefficient.

After the derivation and calculation, we can derive the flux attenuation formula for the overdense case as
\begin{equation}
    P_{\rm R} =  P_{\rm T}G_{\rm R}\frac{\lambda^2 }{4\pi (2 R \sin^2 \theta_q + r \sin \theta_q \cos \varphi_q)} r \cos \phi .
    \label{overdence calcu}
\end{equation}

By replacing the specular reflection process with Thompson scattering of the individual particles and integrating the particles in the trail, one can also derive the flux decay formula for the underdense case as
\begin{equation}
    P^{'}_{\rm R} =  P_{\rm T}G_{\rm R}\frac{\lambda^3 q^2 r_{\rm e}^{2}}{4\pi R  \sin^2 \theta_q } \cdot \exp\left( \frac{8\pi ^2r_{0}^{2}}{\lambda ^2\sec ^2\phi} \right)\exp{\left( \frac{-32\pi ^2Dt}{\lambda ^2\sec ^2\phi} \right)} . \label{underdence calcu}
\end{equation}

To obtain the final reception equation, it is also necessary to consider the equivalent metallic cylinder radius of the meteor trail as (\citealt{14403})
\begin{equation}
r_{\rm c}=\left[ 4Dt\cdot \ln \left( \frac{r _e q\lambda ^2\sec ^2\phi}{4\pi ^2Dt} \right) \right] ^{\frac{1}{2}} .
\end{equation}
Together, we can get
\begin{equation}
\begin{split}
    P_{\rm R} & = P_{\rm T}G_{\rm R}\frac{\lambda ^2}{4\pi (2 R \sin^2 \theta_q+r \sin \theta_q \cos \varphi_q)} \\ & \times \left[ \frac{4Dt}{\sec ^2\phi}\cdot \ln \left( \frac{r_{\rm e} q\lambda ^2\sec ^2\phi}{4\pi ^2Dt} \right) \right] ^{\frac{1}{2}} .
\end{split}
\end{equation}
After the elapsed time approximates
\begin{equation}
    \tau '=\frac{r_{\rm e}q\lambda ^2\sec ^2\phi}{4\pi ^2D} .
\end{equation}
the trail becomes an underdense trail.

The derivative of $P_{\rm R}$ with respect to $r$ in the equation shows that its derivative is greater than $0$ in all the ranges of values we studied, so $P_{\rm R}$ increases monotonically with respect to $r$. The derivative of $r$ with respect to time $t$ shows that $r$ is maximum when $ t = \tau '/e$ where
\begin{equation}
    r\rightarrow \left( \frac{r_{\rm e}q\lambda ^2 \sec ^2\phi}{\pi ^2e} \right) ^{\frac{1}{2}} .
\end{equation}
here $e$ is the base of the natural logarithm. From this, we can obtain the maximum received power with the value
\begin{align}
    P_{\rm R} & = P_{\rm T}G_{\rm R}\frac{\lambda ^3}{4\pi ^2\left( 2 R \sin^2 \theta_q +\left( \frac{r_{\rm e}q\lambda ^2 \sec ^2\phi}{\pi ^2e} \right)^{\frac{1}{2}}  \sin \theta_q \cos \varphi_q \right)}\left( \frac{r_{\rm e}q}{e} \right) ^{\frac{1}{2}} \nonumber \\
    & = P_{\rm T} G_{\rm R}\frac{\lambda ^3}{4\pi^2 \left( 2 R \sin^2 \theta_q \left(\frac{e}{r_{\rm e} q}  \right)^{\frac{1}{2}}+\left( \frac{\lambda \sec \phi}{\pi} \right) \sin \theta_q \cos \varphi_q \right)} .
\end{align}
where the second term of the denominator is very small with respect to the first term and can be neglected in the calculation.

\bsp	
\label{lastpage}
\end{document}